\documentclass[manuscript,screen,nonacm]{acmart}

\AtBeginDocument{%
  \providecommand\BibTeX{{Bib\TeX}}}

\setcopyright{none}
\renewcommand\footnotetextcopyrightpermission[1]{}
\acmConference{}{}{}
\acmDOI{}
\acmISBN{}

\begin{document}

\title[Gendered Double Standards in CelebA]{Gendered Double Standards in CelebA: A Multidimensional Audit}

\author{Sieun Park}
\affiliation{%
  \institution{London School of Economics and Political Science}
  \department{Department of Methodology}
  \city{London}
  \country{United Kingdom}
}
\email{sieun5393@gmail.com}

\author{Yuanmo He}
\orcid{0000-0001-7827-5395}   % optional
\affiliation{%
  \institution{London School of Economics and Political Science}
  \department{Department of Methodology}
  \city{London}
  \country{United Kingdom}
}
\email{yuanmohe@gmail.com}

\authorsaddresses{Authors' Contact Information: Sieun Park, sieun5393@gmail.com; Yuanmo He, yuanmohe@gmail.com. This work was conducted while both authors were at the Department of Methodology, London School of Economics and Political Science, United Kingdom.}

\begin{abstract}
Media scholarship has long documented gendered double standards of ageing and beauty. Yet how these standards appear in combinations of visual attributes and their evaluation in computer vision datasets remains underexplored. We examine how gendered appearance norms are reflected in CelebA's subjective ``Attractive'' annotation and in models trained to predict it. Drawing on a multidimensional approach to representational bias, we audit 202,599 images across 1) dataset structure, 2) learned feature associations, and 3) spatial model sensitivity. Hierarchical clustering identifies seven attribute bundles, including groupings associated with youth, makeup, and adornment, and with ageing, facial hair, and formal attire. Logistic regression with cluster-by-perceived-sex interactions reveals an asymmetry obscured by overall label rates. Although female-perceived faces receive positive attractiveness annotations more frequently overall, an ageing- and masculine-coded bundle has a substantially stronger negative association with their attractiveness odds than with those of male-perceived faces, relative to a youthful, feminine-coded reference bundle. SHAP analysis of XGBoost predictions identifies related differences: body-size attributes contribute negatively to predictions for female-perceived faces but are approximately neutral or slightly positive for male-perceived faces. Exploratory Grad-CAM analysis contrasts central facial sensitivity among the highest-scoring young female-perceived faces with predominantly upper-forehead sensitivity among older male-perceived faces. Taken together, these findings provide complementary evidence consistent with gendered appearance double standards, particularly a narrower association of female attractiveness annotations with youthful, feminine-coded presentation. The study demonstrates how theoretically grounded, multidimensional audits can reveal evaluative patterns that overall label rates and predictive accuracy leave unresolved.
\end{abstract}




\maketitle

\section{Introduction}
\label{sec:introduction}

Images are a powerful mode of human communication. Visual content is more memorable than verbal information \citep{grady1998}, evokes stronger affective responses than text \citep{brader2005}, and generates higher engagement in social and political settings \citep{rogers2014}. By presenting associations simultaneously, images condense meaning into seemingly self-evident cues. Cultural conventions can thus appear as objective representation \citep{barthes1972,hall1997}, shaping perceptions of identity, value, and social worth \citep{berger1972}.

Media images are central to these visual codes. As \citet{goffman1976} argued, such ``public pictures'' express and reproduce gendered hierarchies and cultural values. Advertising, film, and news photography have often associated femininity with youth, beauty, and ornamentation, and masculinity with authority, maturity, and expertise \citep{goffman1979,sontag1972,wolf1991}. Visual culture also reproduces racial hierarchies by valorising lighter skin while marginalising darker complexions \citep{hunter2007,dyer1997}. These recurring conventions shape public understandings of gender and race.

These conventions can enter computational vision through facial datasets drawn from celebrity photography, editorial portraits, and platform-curated media \citep{yazdani2022distraction}. Appearance norms concerning who is represented, how they are styled, and which traits are highlighted can shape both image collections and annotation frameworks \citep{bartl2023}. Models trained on these datasets may consequently learn associations reflecting cultural hierarchies of recognition and evaluation \citep{steed2020}.

In facial analysis, demographic audits have documented
classification disparities and motivated efforts to reduce
them \citep{buolamwini2018,raji2019}. Research on
representational harms further examines which labels systems
apply to different groups and the social meanings those labels
convey \citep{crawford2017,katzman2023}. For example,
\citet{schwemmer2020} show that image recognition systems
describe women more frequently through appearance, while
\citet{katzman2023} discuss the unequal application of beauty
labels. These findings establish that the distribution and
content of labels can reinforce gendered representations.
They leave a further question, however: when a favourable
appearance label is applied, are the appearance traits
associated with receiving it different across groups?

This distinction matters because more frequent positive
evaluation need not imply less restrictive appearance
expectations. Media scholarship describes a gendered double
standard in which women's beauty is more closely tied to
youth and feminine presentation, while ageing remains more
compatible with valued representations of men
\citep{sontag1972,wolf1991}. An audit of overall beauty-label
rates could therefore find greater positivity towards women
while leaving this asymmetry unexamined. Similarly, a model
could accurately reproduce the labels while also reproducing
their gendered appearance associations. Evaluating such a
model requires examining the judgments encoded in its target,
rather than predictive performance alone
\citep{jacobs2021measurement}.

We address this question in CelebA, a widely used facial-analysis
benchmark whose binary ``Attractive'' annotation records
subjective judgments \citep{liu2015}. We investigate gendered
appearance double standards as one manifestation of
representational bias, examining whether higher overall
positive-label rates for female-perceived faces coexist with
stronger negative associations for ageing- or masculine-coded
appearance bundles relative to a youthful, feminine-coded
bundle. We then examine where related associations recur in
models trained to predict the annotation. Our aim is to connect
documented gendered appearance norms to measurable patterns
within a benchmark's evaluative labels, treating those labels
as judgments to investigate rather than objective measures
of attractiveness.

The audit proceeds at three levels. First, hierarchical
clustering identifies co-occurring ``trait bundles,'' and
cluster-by-perceived-sex interactions examine their
associations with attractiveness annotations. Second,
attribute-based classifiers and feature attribution analysis
examine learned appearance associations. Third, an image-based
classifier and exploratory Grad-CAM analysis describe spatial
sensitivity across perceived-sex and ``Young'' subgroups,
alongside accuracy and average precision.

Our contribution is to distinguish how frequently groups
receive favourable evaluation from the appearance patterns
associated with that evaluation. The findings show how higher
overall positivity can coexist with gendered evaluative
associations, some of which recur in trained models. This
extends audits of label distributions and descriptive content
by examining the appearance associations within a shared favourable category.

\section{Related Works}
\label{sec:literature-review}

\subsection{Gender Double Standards in Media}
\label{sec:gender-double-standards}

Media scholarship has long shown that appearance is evaluated through gendered double standards. From news photography to advertising, women and men are visually coded in systematically different ways that reinforce hierarchies of social value \citep{becker1974,ferree1990}. \citet{goffman1976,goffman1979} described advertisements as public pictures where gendered structures of hierarchy are repeatedly staged through what he called ``gender displays.'' These visual conventions tie authority to masculinity, signalled by suits, upright posture, or facial hair, while femininity is equated with youth and ornamentation through makeup, styled hair, and bodily display \citep{goffman1976,kang1997}. \citet{wolf1991} shows the same pattern in broadcast news. Male anchors are judged primarily on expertise and can enhance their authority through visible signs of maturity, while female anchors are evaluated above all on youth and appearance. Women face a narrow, rigid set of beauty requirements, whereas men enjoy a wide latitude of acceptable looks. As Wolf observes, ``If a single standard were applied equally to men as to women in TV journalism, most of the men would be unemployed'' \citep{wolf1991}.

This visual norm has not disappeared with the shift to digital platforms. Several media studies show that these double norms have been adapted and amplified. Research on Instagram selfies shows that Goffman's classic ``gender displays'' continue to structure online self-presentation, with men more often depicted as agentic and self-possessed, and women framed as ornamental or decorative \citep{doering2016}. The rise of the influencer economy further intensifies this pattern by codifying what \citet{duffy2015} describe as ``entrepreneurial femininity.'' This term refers to a carefully curated aesthetic of youthfulness, polish, and consumption that resonates with audience expectations while also aligning with platform algorithms \citep{cotter2019}. These dynamics contribute to persistent visibility gaps, particularly along lines of gender and age. \citet{edstrom2018} shows older women remain markedly underrepresented across digital channels compared to their male counterparts. \citet{smith2018} draw a similar pattern in content analyses in the film industry, identifying major male characters who concentrate in their 30s--40s and persist on screen later in life, while major female characters cluster in their 20s--30s and drop off with age. \citet{clarke2008} showed that older women engage in intensive ``beauty work'' to resist ageist penalties in workplaces and intimate spheres. Such research consistently shows that gender, age, and beauty are not simply individual traits but cultural codes that determine who is seen, valued, and granted visibility across media domains. These codes endure across both legacy and digital platforms, embedding a double standard.

\subsection{Algorithmic Bias in Computer Vision}
\label{sec:algorithmic-bias}

Contemporary vision datasets draw on curated media sources, such as celebrity photography or editorial images, which may not reflect demographic reality \citep{yazdani2022distraction}. These sources shape who is visible and how groups are portrayed, while annotation taxonomies encode assumptions about which attributes matter \citep{fabbrizzi2021,scheuerman2021}. Cultural norms can therefore become embedded in both images and labels, creating conditions for their reproduction in model behaviour.

A substantial strand of algorithmic fairness research addresses disparities in outcomes and error rates across demographic groups \citep{barocas2019}. Technical approaches include data pre-processing \citep{kamiran2012}, learning fair representations \citep{zemel2013,madras2018}, and prediction adjustments to satisfy fairness criteria \citep{hardt2016}. In facial analysis, external audits and subsequent system revisions have reduced error rates for darker-skinned women \citep{buolamwini2018,raji2019}. Related disparities have been examined in attribute classification \citep{das2018eccvw,denton2019counterfactual,karkkainen2021fairface}, expression recognition \citep{chen2021ferbias,xu2020investigating}, and face recognition and verification \citep{gong2020joint,robinson2020toobias}.

Performance improvements alone do not address representational harms arising from how people are depicted, labelled, and evaluated \citep{crawford2017,katzman2023}. Existing studies examine harmful categories and collection practices \citep{crawford2019,prabhu2020}, amplified gender associations \citep{zhao2017,hendricks2018womenalsosnowboard}, stereotypical question--answer content \citep{hirota2022}, and beauty labels \citep{katzman2023}. Other work identifies the disproportionate description of women through appearance \citep{schwemmer2020} and associations in CelebA that performance disparities alone do not capture \citep{serianni2025}. Complementary research in language models identifies stereotypical associations in word embeddings \citep{caliskan2017}. More recently, \citet{bai2025explicitly} show that aligned language models can exhibit stereotypical associations and discriminatory choices despite appearing unbiased on conventional evaluations. Together, these findings support examining the content of learned associations alongside aggregate outcomes.

For subjective appearance labels, a further question concerns whether the same label encodes different evaluative standards across groups. A higher rate of Attractive'' annotation does not establish that a group faces less restrictive appearance standards: favourable judgment may be more strongly associated with particular traits, while departures from those traits carry stronger negative associations. Drawing on scholarship on gendered appearance norms \citep{goffman1979,sontag1972,wolf1991}, we examine which traits and appearance bundles are associated with Attractive'' within each perceived-sex category. Specifically, we investigate whether higher overall positivity among female-perceived faces coexists with stronger negative associations for ageing- or masculine-coded appearance bundles, and whether related patterns recur in models trained to predict the annotation. This approach examines gendered evaluative double standards within a shared benchmark label.

\section{Methodology}
\label{sec:methodology}

\subsection{Dataset}
\label{sec:dataset}

The CelebFaces Attributes \citep{liu2015} dataset contains 202,599 celebrity images collected from online media sources. Because its subjects are public figures, CelebA directly reflects the conventions of media representation that structure visibility and desirability. It provides 40 binary attributes spanning morphological traits (e.g., high cheekbones, big nose), cosmetic cues (e.g., heavy makeup, lipstick), and age-related indicators (e.g., gray hair), alongside labels for ``Attractive'' and ``Male.'' This research does not consider ``Attractive'' objective ground truth, but the product of annotators' subjective and socially constructed judgments.

CelebA has been used extensively for facial attribute classification \citep{liu2015,rudd2016}, face recognition \citep{deng2019,guo2020,zhang2016}, and generative facial editing using Generative Adversarial Networks \citep{choi2018,he2019,shen2020,wu2019} and Variational Autoencoders \citep{higgins2017,yan2016}. More recently, it has become central to fairness research for identifying biases \citep{conti2022,zhang2018} and developing debiasing methods \citep{yang2022,shrestha2021,zhangsang2020}. In this project, CelebA is examined as a site where cultural hierarchies of gender, age, and appearance may be statistically encoded. Rather than infer bias from any single disparity, we investigate whether theoretically anticipated gendered appearance norms recur across attribute and label structure, learned feature associations, and spatial model sensitivity.

% TODO: Add the intended bibliography entry for wang2022.
\textbf{Gender and Sex Labels.} This study uses female (\texttt{Male} = 0) and male (\texttt{Male} = 1) as shorthand for the two categories defined by CelebA's binary ``Male'' attribute. These labels reflect how subjects were externally classified from visual appearance, rather than biological sex or self-identified gender. We therefore interpret them as perceived binary categories, consistent with prior work using externally assigned demographic annotations \citep{wang2022}. CelebA's documentation also provides limited information about who assigned the labels, how ambiguous cases were resolved, or how consistently the categories were applied. Accordingly, our findings are limited to patterns conditioned on CelebA's externally assigned binary labels; they do not address participants' gender identities, non-binary identities, or biological sex. Future work using self-identified gender information could move beyond this binary, perception-based framework.

\subsection{Trait Clustering and Label Analysis}
\label{sec:trait-clustering}

We first identify clusters of co-occurring appearance traits in CelebA and interpret their composition in relation to cultural norms of gender and appearance. We then examine how these trait bundles relate to the dataset's attractiveness annotations, testing whether the appearance profiles associated with positive evaluations differ between female-perceived and male-perceived faces. 

We compute a Pearson correlation matrix across CelebA's 39 non-target attributes, excluding ``Attractive.'' The resulting \(39 \times 39\) matrix describes pairwise associations between attributes across the 202,599 images. Correlations are transformed into dissimilarities:
\begin{equation}
d_{ab} = \frac{1-r_{ab}}{2},
\label{eq:1}
\end{equation}
where \(r_{ab}\) is the Pearson correlation between attributes \(a\) and \(b\). This transformation maps perfect positive correlation to 0, zero correlation to 0.5, and perfect negative correlation to 1.

We apply hierarchical agglomerative clustering (HAC) with average linkage to the dissimilarity matrix and reorder the correlation heatmap by the resulting dendrogram. An elbow analysis of mean intra-cluster dissimilarity across \(K=2\)--\(9\) motivates seven clusters as a practical balance between compactness and interpretability (Figure~\ref{fig:elbow}). A t-SNE projection of the same matrix provides a complementary visualisation of local attribute similarities \citep{vandermaaten2008}. Further details are provided in the Appendix.

To relate attribute clusters to image-level annotations, each image is assigned to the cluster with the highest coverage ratio:
\begin{equation}
\mathrm{ratio}_{ik}
=
\frac{1}{|S_k|}
\sum_{a\in S_k} x_{ia},
\label{eq:2}
\end{equation}
where \(S_k\) is the set of attributes in cluster \(k\), \(|S_k|\) is its size, and \(x_{ia}=1\) when attribute \(a\) is present in image \(i\), and 0 otherwise. Normalising by cluster size allows comparison across bundles containing different numbers of attributes. This assignment identifies the bundle with the greatest proportional coverage in each image; images may still contain attributes from several bundles.

We then construct image-level records containing cluster assignment, perceived-sex category, and attractiveness annotation. For each cluster--perceived-sex pair, we calculate the proportion of images labelled ``Attractive'' and estimate approximate 95\% confidence intervals using the binomial standard error.

To examine whether cluster associations differ across perceived sex, we fit a logistic regression with cluster indicators and interaction terms:
\begin{equation}
\begin{split}
\log\!\left(\frac{p_i}{1-p_i}\right)
={}& \beta_0
+\sum_{k=2}^{K}\beta_k C_{ik}
+\gamma M_i \\
&+\sum_{k=2}^{K}\delta_k(C_{ik}M_i),
\end{split}
\label{eq:3}
\end{equation}
where \(p_i=P(\mathrm{Attractive}_i=1)\), \(C_{ik}\) indicates membership in cluster \(k\), and \(M_i\) corresponds to CelebA's binary ``Male'' annotation. The reference cluster and female-perceived category (\(M_i=0\)) define the baseline. The coefficients \(\beta_k\) represent cluster differences within the female-perceived group, while \(\gamma\) represents the male--female difference within the reference cluster. Each interaction coefficient \(\delta_k\) indicates how the corresponding cluster association differs between male-perceived and female-perceived images.

\subsection{Predictive Modeling and Feature Attribution}
\label{sec:predictive-modeling}

To examine whether dataset-level appearance associations recur in learned predictions, we train a pooled XGBoost classifier \citep{chen2016xgboost} to predict ``Attractive'' from the 39 non-target attributes, including ``Male.'' We also fit separate classifiers for female-perceived and male-perceived faces using the 38 appearance attributes, excluding ``Male.'' XGBoost supports non-linear interactions among co-occurring attributes and efficient TreeSHAP attribution.

Hyperparameters are tuned through 80 Bayesian optimisation trials using Optuna \citep{akiba2019}, covering tree depth, learning rate, boosting rounds, minimum child weight, observation and feature subsampling, and regularisation. We retain the configuration with the highest cross-validated ROC-AUC. Training uses histogram-based tree construction on an NVIDIA A100 GPU.

We apply SHAP \citep{Lundberg2017} using TreeSHAP \citep{lundberg2020} to attribute predictions to individual features. Test-set attributions are aggregated to obtain global feature rankings and examine the direction of contributions in the pooled and subgroup models. Because attributes co-occur and subgroup models are fitted separately, attributions depend on the fitted model, feature-dependence assumptions, and each model's reference prediction. We interpret them as learned predictive associations rather than independent causal effects of appearance traits.

\subsection{Saliency Visualisation and Subgroup Outcomes}
\label{sec:saliency}

Finally, we examine whether the gendered asymmetries identified in the preceding analyses are also reflected in spatial sensitivity and subgroup performance. We use an ImageNet-pretrained ResNet-18 \citep{he2016}, selected as a standard convolutional architecture that supports spatial localisation with Grad-CAM while remaining computationally tractable for image-level auditing.

All CelebA images are resized to \(224 \times 224\) pixels and normalised before being passed to the network. Following the transfer-learning approach examined by \citet{yosinski2014}, all layers except the final fully connected classifier are frozen, disabling gradient computation for their parameters. This approach assumes that features learned on the source task remain informative for the target task, allowing adaptation through the task-specific classifier.

The classifier is trained for 10 epochs using batches of 64 images for optimisation and 128 images for validation and testing. We use Adam \citep{kingma2015}, rather than the SGD optimiser with momentum of 0.9 used in the original ResNet training procedure \citep{he2016}. The initial learning rate is 0.01, with exponential decay applied after each epoch by multiplying the learning rate by 0.9. This schedule progressively reduces the optimisation step size during training.

To examine spatial sensitivity, we apply Grad-CAM to the final convolutional layer, generating saliency maps that highlight regions associated with the model's predictions \citep{Selvaraju_2017_ICCV}. We define four subgroups by crossing CelebA's binary ``Young'' and ``Male'' annotations: young female (\(\mathrm{Young}=1,\mathrm{Male}=0\)), young male (\(\mathrm{Young}=1,\mathrm{Male}=1\)), older female (\(\mathrm{Young}=0,\mathrm{Male}=0\)), and older male (\(\mathrm{Young}=0,\mathrm{Male}=1\)). These are externally assigned binary annotations, not verified ages, biological characteristics, or self-identified gender identities.

For each subgroup, we select the Top-\(K\) images with the highest predicted probabilities for the positive ``Attractive'' class (\(K=64\)). Grad-CAM maps are computed for each selected image and then averaged to produce subgroup-level overlays on group-average faces. These visualisations support qualitative comparisons of spatial sensitivity among the highest-scoring images in each subgroup.

In addition to qualitative comparisons, we report subgroup performance using accuracy and average precision (AP). Accuracy measures the proportion of correct predictions at a fixed threshold, whereas AP summarises precision across recall levels as the threshold varies. Reporting both provides complementary views of classification and ranking performance, although AP remains sensitive to differences in positive-label prevalence across subgroups.

\section{Results}

\subsection{Trait Clustering and Label Analysis}

\subsubsection{Latent Attribute Structure in CelebA}

\begin{figure*}[t]
\centering
\includegraphics[width=0.95\textwidth]{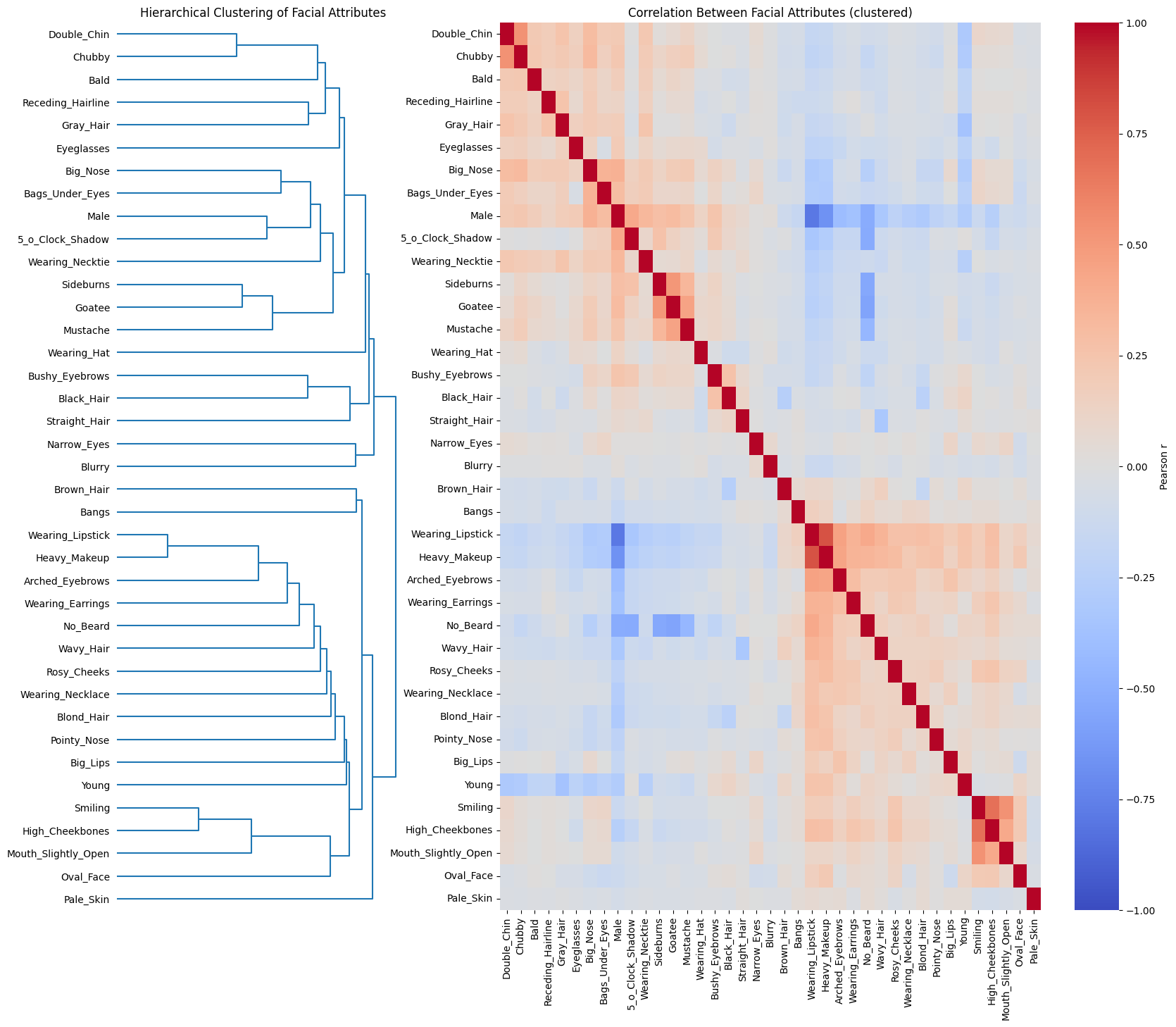}
\caption{Pearson correlations among the 39 CelebA attributes excluding ``Attractive,'' reordered according to the average-linkage hierarchical clustering. The dendrogram displays the nested relationships underlying the seven-cluster solution.}
\label{fig:heatmap}
\end{figure*}

The two largest attribute clusters organise distinct combinations of appearance cues. Cluster~1 combines cosmetic and expression-related attributes with ``Young,'' while Cluster~6 brings together facial-hair, ageing, and body-size annotations alongside ``Male'' and ``Wearing Necktie'' (Table~\ref{tab:cluster_membership}; Figure~\ref{fig:heatmap}). The remaining clusters contain smaller groups of generally weakly associated attributes or single annotations. A complementary t-SNE projection is provided in the Appendix.

\begin{table}[t]
\centering
\caption{Attribute membership in the seven-cluster solution obtained using average-linkage hierarchical agglomerative clustering.}
\label{tab:cluster_membership}
\footnotesize
\setlength{\tabcolsep}{4pt}
\begin{tabular}{@{}p{0.15\columnwidth}p{0.80\columnwidth}@{}}
\toprule
\textbf{Cluster} & \textbf{Attributes} \\
\midrule
Cluster 1 & Arched Eyebrows, Big Lips, Blond Hair, Rosy Cheeks, Mouth Slightly Open, No Beard, High Cheekbones, Heavy Makeup, Oval Face, Pointy Nose, Smiling, Young, Wearing Necklace, Wearing Lipstick, Wearing Earrings, Wavy Hair \\
\addlinespace[2pt]
Cluster 2 & Brown Hair, Bangs \\
\addlinespace[2pt]
Cluster 3 & Pale Skin \\
\addlinespace[2pt]
Cluster 4 & Narrow Eyes, Blurry \\
\addlinespace[2pt]
Cluster 5 & Black Hair, Straight Hair, Bushy Eyebrows \\
\addlinespace[2pt]
Cluster 6 & Goatee, Eyeglasses, Double Chin, Chubby, Bald, Bags Under Eyes, 5 o'Clock Shadow, Big Nose, Sideburns, Receding Hairline, Male, Mustache, Gray Hair, Wearing Necktie \\
\addlinespace[2pt]
Cluster 7 & Wearing Hat \\
\bottomrule
\end{tabular}
\end{table}

\begin{figure}[t]
\centering
\includegraphics[width=0.5\columnwidth,keepaspectratio]{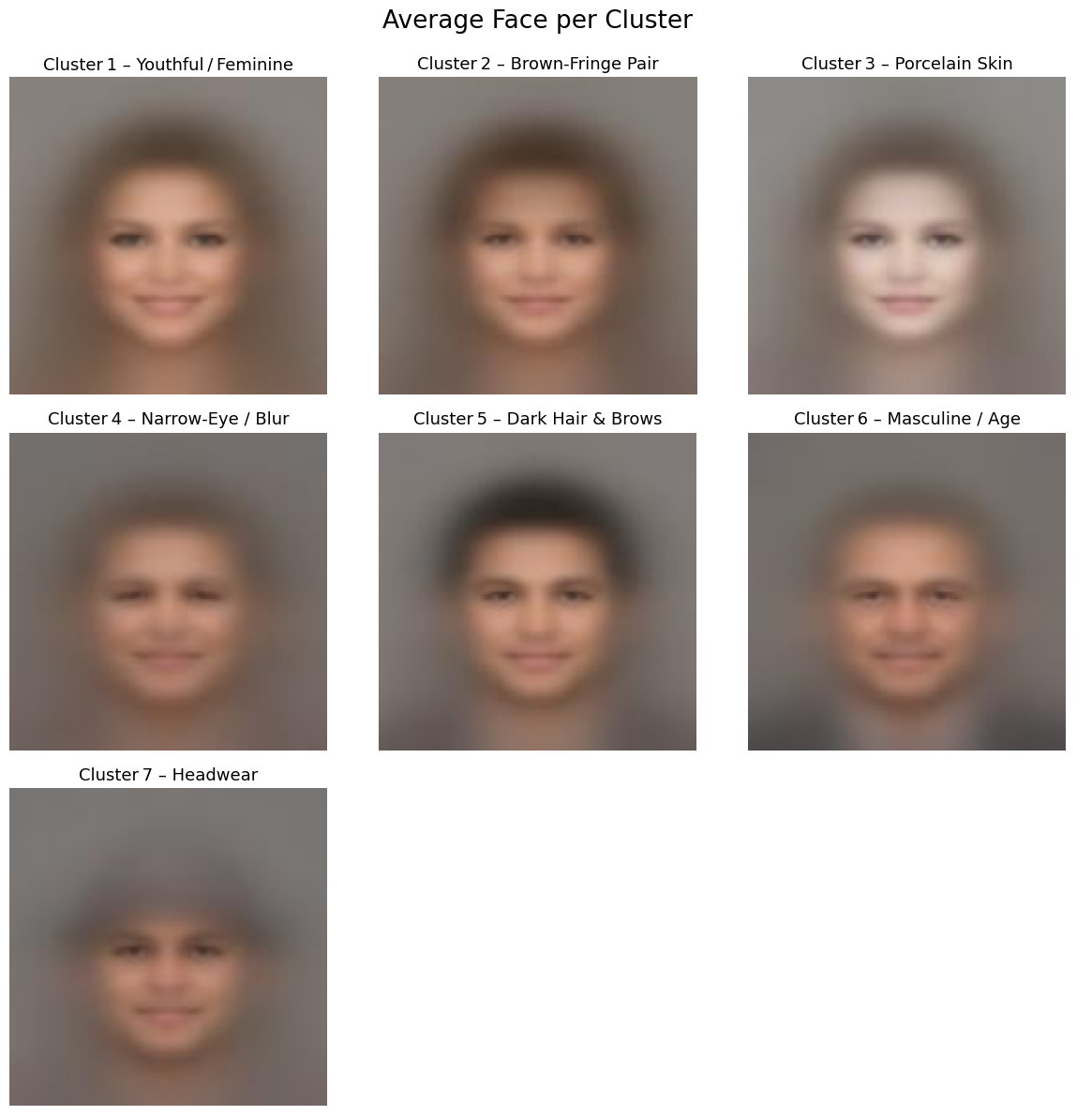}
\caption{Average facial composites for images assigned to the seven attribute clusters using the coverage-ratio procedure. Cluster identifiers correspond to the attribute bundles in Table~\ref{tab:cluster_membership}.}
\label{fig:clusters}
\end{figure}

Cluster~1 brings together cosmetic, expression-related, and
youth annotations. Heavy Makeup and Wearing Lipstick have
the strongest positive off-diagonal correlation in the matrix
(\(r=0.802\)) and are strongly negatively associated with
``Male'' (\(r=-0.667\) and \(r=-0.789\), respectively).
Smiling and High Cheekbones form an expression-related
subgroup (\(r=0.683\)). Young is modestly correlated with
Heavy Makeup and Wearing Lipstick (\(r=0.243\) and
\(r=0.252\)), but approximately uncorrelated with Smiling
and High Cheekbones. The bundle thus connects several
appearance dimensions rather than a uniformly correlated
set of traits.

Cluster~6 combines body-size, ageing, facial-hair, and
formalwear annotations. Associations are stronger within
subgroups, such as Double Chin and Chubby (\(r=0.534\))
and Goatee and Sideburns (\(r=0.513\)), than between them.
For example, the body-size annotations have weaker
associations with Bald, Receding Hairline, and Gray Hair
(\(r=0.187\)--\(0.253\)). Wearing Necktie is positively
associated with ``Male'' (\(r=0.330\)), but only modestly
with facial-hair annotations (\(r=0.060\)--\(0.103\)).
The cluster therefore links grooming, ageing, and clothing
cues without directly measuring professional status or
authority.

The smaller clusters contain generally weaker associations:
Brown Hair and Bangs form Cluster~2 (\(r=0.069\)),
Narrow Eyes and Blurry form Cluster~4 (\(r=0.071\)),
and Black Hair, Straight Hair, and Bushy Eyebrows form
Cluster~5 (\(r=0.074\)--\(0.258\)). Pale Skin and Wearing
Hat remain singleton clusters. These groupings alone do
not establish a shared cultural or photographic explanation;
in particular, Pale Skin is not a validated measure of
skin tone or racial identity.

Together, these findings describe the joint organisation
of appearance and perceived-sex annotations, including
``Male'' as a clustering input. We next examine whether
the bundles' associations with ``Attractive'' differ
across perceived sex.

\subsubsection{Gendered Beauty Penalties}

\begin{figure}[t]
  \centering
    \includegraphics[width=0.8\columnwidth]{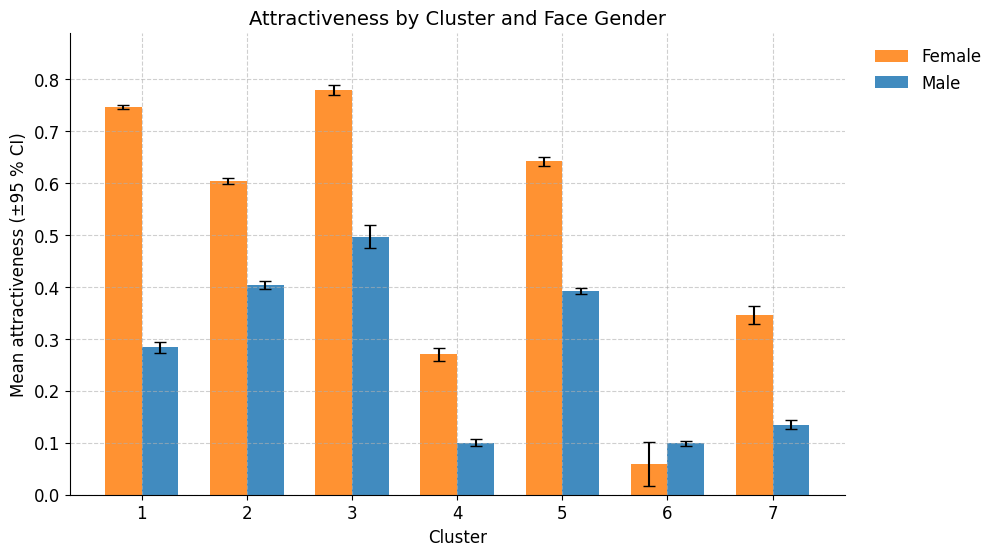}
  \caption{Proportion of images annotated as ``Attractive'' across the seven attribute clusters, stratified by perceived sex. Error bars indicate approximate 95\% confidence intervals based on the binomial standard error. Female-perceived faces have higher positive-label proportions in all clusters except Cluster~6.}
  \label{fig:cluster_mean}
\end{figure}

Female-perceived faces receive a higher proportion of ``Attractive'' annotations in every cluster except Cluster~6 (Figure~\ref{fig:cluster_mean}). However, this greater positivity coexists with stronger negative associations for several appearance bundles relative to the youthful, feminine-coded reference cluster (Cluster 1). Clusters~4, 6, and~7 have more negative contrasts for female-perceived faces, while Clusters~2 and~5 show negative contrasts for female-perceived faces but positive contrasts for male-perceived faces. These patterns distinguish how frequently groups receive positive annotations from the appearance conditions associated with those evaluations.

The logistic regression estimates cluster contrasts relative to Cluster~1 separately across perceived-sex groups (Table~\ref{tab:logit_full}). Within the reference cluster, the estimated log odds are \(1.081\) for female-perceived faces and \(1.081-2.008=-0.927\) for male-perceived faces. These correspond to positive-label probabilities of approximately \(0.747\) and \(0.284\), respectively. Cluster~1 serves as the comparison category because it combines youthful, cosmetic, and expression-related attributes; its selection as the reference does not establish it as an objective standard of attractiveness.

\begin{table}[t]
    \centering
    \caption{Logistic regression predicting the ``Attractive'' annotation. The reference category is Cluster~1 among female-perceived faces. Coefficients are reported with standard errors in parentheses. Exponentiated coefficients and their 95\% confidence intervals appear in the final column: the intercept represents baseline odds, interaction terms represent ratios of odds ratios, and other terms represent odds ratios. All individual coefficients have \(p<0.001\).}
    \label{tab:logit_full}
    \footnotesize
    \setlength{\tabcolsep}{6pt}
    \begin{tabular}{lcc}
    \toprule
    Predictor & Coef. (SE) & Exp(coef.) [95\% CI] \\
    \midrule
    Intercept          &  1.0812\ (0.009) &  2.95\ [2.90,\ 3.00]  \\
    Cluster 2          & -0.6604\ (0.016) &  0.52\ [0.50,\ 0.53]  \\
    Cluster 3          &  0.1802\ (0.031) &  1.20\ [1.13,\ 1.27]  \\
    Cluster 4          & -2.0757\ (0.031) &  0.13\ [0.12,\ 0.13]  \\
    Cluster 5          & -0.4981\ (0.022) &  0.61\ [0.58,\ 0.63]  \\
    Cluster 6          & -3.8538\ (0.390) &  0.02\ [0.01,\ 0.05]  \\
    Cluster 7          & -1.7171\ (0.041) &  0.18\ [0.17,\ 0.19]  \\
    Male               & -2.0081\ (0.026) &  0.13\ [0.13,\ 0.14]  \\
    Cluster 2 $\times$ Male & 1.2003\ (0.034) & 3.32\ [3.11,\ 3.55] \\
    Cluster 3 $\times$ Male & 0.7341\ (0.061) & 2.08\ [1.85,\ 2.35] \\
    Cluster 4 $\times$ Male & 0.8107\ (0.056) & 2.25\ [2.02,\ 2.51] \\
    Cluster 5 $\times$ Male & 0.9903\ (0.035) & 2.69\ [2.51,\ 2.88] \\
    Cluster 6 $\times$ Male & 2.5713\ (0.391) & 13.08\ [6.07,\ 28.16] \\
    Cluster 7 $\times$ Male & 0.7875\ (0.060) & 2.20\ [1.95,\ 2.47] \\
    \midrule
    \multicolumn{3}{l}{\(N=202{,}599\); McFadden's pseudo-\(R^{2}=0.175\).} \\
    \multicolumn{3}{l}{Log-likelihood \(\approx-115{,}800\); likelihood-ratio test \(p<0.001\).} \\
    \bottomrule
    \end{tabular}
\end{table}

Cluster~2, comprising Brown Hair and Bangs, has a negative contrast for female-perceived faces (\(-0.660\)). For male-perceived faces, the positive interaction coefficient (\(1.200\)) produces a net cluster contrast of \(0.540\). Cluster~5, comprising Black Hair, Straight Hair, and Bushy Eyebrows, shows the same directional reversal: its contrast is \(-0.498\) for female-perceived faces and \(0.492\) for male-perceived faces. These positive male-perceived contrasts concern comparisons with male-perceived faces in Cluster~1; they do not indicate higher positive-label probabilities than those of female-perceived faces in the corresponding clusters.

The largest negative contrast occurs for Cluster~6, which combines facial-hair, ageing, and body-size annotations with ``Male'' and Wearing Necktie. Relative to Cluster~1, its coefficient is \(-3.854\) for female-perceived faces, compared with a net contrast of \(-1.283\) for male-perceived faces. The corresponding within-group odds ratios are approximately \(0.021\) and \(0.277\), indicating a greater relative reduction in attractiveness odds for female-perceived faces. Estimated positive-label probabilities fall from \(0.747\) in Cluster~1 to \(0.059\) in Cluster~6 for female-perceived faces, and from \(0.284\) to \(0.099\) for male-perceived faces. Cluster~6 is therefore the only cluster in which the ordering of positive-label proportions reverses. Its interaction estimate is also less precise than the other interactions, with an exponentiated coefficient of \(13.08\) and a 95\% confidence interval of \([6.07,28.16]\).

Clusters~4 and~7 have negative contrasts for both groups, with larger reductions on the log-odds scale for female-perceived faces. For Cluster~4, comprising Narrow Eyes and Blurry, the contrasts are \(-2.076\) for female-perceived faces and \(-1.265\) for male-perceived faces. For Cluster~7, defined by Wearing Hat, the corresponding contrasts are \(-1.717\) and \(-0.930\). These results establish differences in cluster associations without identifying why the annotations differ.

Cluster~3, defined by Pale Skin, is the only cluster with a positive contrast relative to Cluster~1 for both groups. Its contrast is \(0.180\) for female-perceived faces and \(0.914\) for male-perceived faces, indicating a stronger positive association for male-perceived faces on the log-odds scale. Nevertheless, estimated positive-label probabilities remain higher for female-perceived faces (\(0.779\)) than for male-perceived faces (\(0.497\)). This finding concerns the dataset's Pale Skin annotation and does not, by itself, establish colorism or differences between racial or ethnic groups.

Across the seven clusters, positive representation differs both in frequency and in its associations with appearance. Higher positive-label rates for female-perceived faces coexist with stronger negative contrasts for several bundles and opposite directions of association for others. These patterns are consistent with gendered appearance double standards within the dataset's annotations. They do not establish causal effects of individual traits or the intentions of annotators. Further robustness checks are provided in the Appendix.

\subsection{Predictive Modeling and Feature Attribution}

The pooled XGBoost classifier, using all 39 non-target attributes including ``Male,'' achieved a test ROC-AUC of \(0.880\) and accuracy of \(0.79\) across 40,520 images (Table~\ref{tab:xgb_perf}). Separate classifiers were fitted to female-perceived and male-perceived faces using the 38 appearance attributes, excluding ``Male,'' to examine subgroup-specific associations.

\begin{table}[t]
\centering
\caption{Pooled XGBoost test performance. Classes 0 and 1 indicate absence and presence of the ``Attractive'' annotation. ROC-AUC was \(0.880\).}
\label{tab:xgb_perf}
\footnotesize
\begin{tabular}{lcccc}
\toprule
Class & Precision & Recall & F1-score & Support \\
\midrule
0 (Negative) & 0.80 & 0.77 & 0.79 & 19,753 \\
1 (Positive) & 0.79 & 0.82 & 0.80 & 20,767 \\
\midrule
Accuracy & \multicolumn{3}{c}{0.79} & 40,520 \\
Macro avg & 0.80 & 0.79 & 0.79 & 40,520 \\
Weighted avg & 0.80 & 0.79 & 0.79 & 40,520 \\
\bottomrule
\end{tabular}
\end{table}

The pooled SHAP summary (Appendix~\ref{app:shap}, Figure~\ref{fig:shap_global}) shows generally positive contributions from Young, Wearing Lipstick, Heavy Makeup, Smiling, and Oval Face, and negative contributions from Chubby, Double Chin, Eyeglasses, Big Nose, Gray Hair, Receding Hairline, and ``Male.''

Subgroup models reveal contrasting associations (Figure~\ref{fig:shap_gender}). Smiling, Wearing Lipstick, and Oval Face generally contribute positively in the female-perceived model but negatively in the male-perceived model; Heavy Makeup contributes positively in both. Chubby and Receding Hairline show the reverse subgroup pattern, while Double Chin and Eyeglasses have more consistently negative contributions among female-perceived faces. These patterns broadly align with the smaller relative Cluster~6 penalty among male-perceived faces, although cluster contrasts do not isolate individual attributes.

Correspondence is not uniform: Pale Skin has predominantly positive contributions in the female-perceived model and smaller, mixed contributions in the male-perceived model, whereas Cluster~3 has a larger positive regression contrast among male-perceived faces. Cluster assignment and feature attribution therefore capture distinct relationships.

Because subgroup classifiers are fitted separately, their SHAP values are interpreted relative to their respective reference predictions. Differences may reflect both learned relationships and subgroup attribute distributions, rather than independent causal effects of appearance traits.

\subsection{Saliency Visualisation and Subgroup Outcomes}

The ResNet-18 classifier shows different performance patterns across the four age--perceived-sex subgroups (Table~\ref{tab:test_per_group}). Young female-perceived faces attain the highest average precision (AP; \(0.9497\)), whereas older male-perceived faces attain the highest accuracy (\(0.9200\)) but the lowest AP (\(0.3977\)). This divergence illustrates why accuracy alone provides an incomplete account of subgroup performance. AP additionally depends on positive-label prevalence, so its variation across groups cannot be attributed solely to differences in model ranking ability.

\begin{table}[t]
\centering
\caption{ResNet-18 test performance across subgroups defined by CelebA's ``Young'' and ``Male'' annotations. AP denotes average precision for the positive ``Attractive'' class.}
\label{tab:test_per_group}
\footnotesize
\begin{tabular}{lcc}
\toprule
Subgroup & Accuracy & AP \\
\midrule
Young female-perceived & 0.8124 & 0.9497 \\
Young male-perceived   & 0.7721 & 0.8002 \\
Older female-perceived & 0.8027 & 0.8028 \\
Older male-perceived   & 0.9200 & 0.3977 \\
\bottomrule
\end{tabular}
\end{table}

\begin{figure}[t]
  \centering
  \includegraphics[width=0.7\columnwidth]{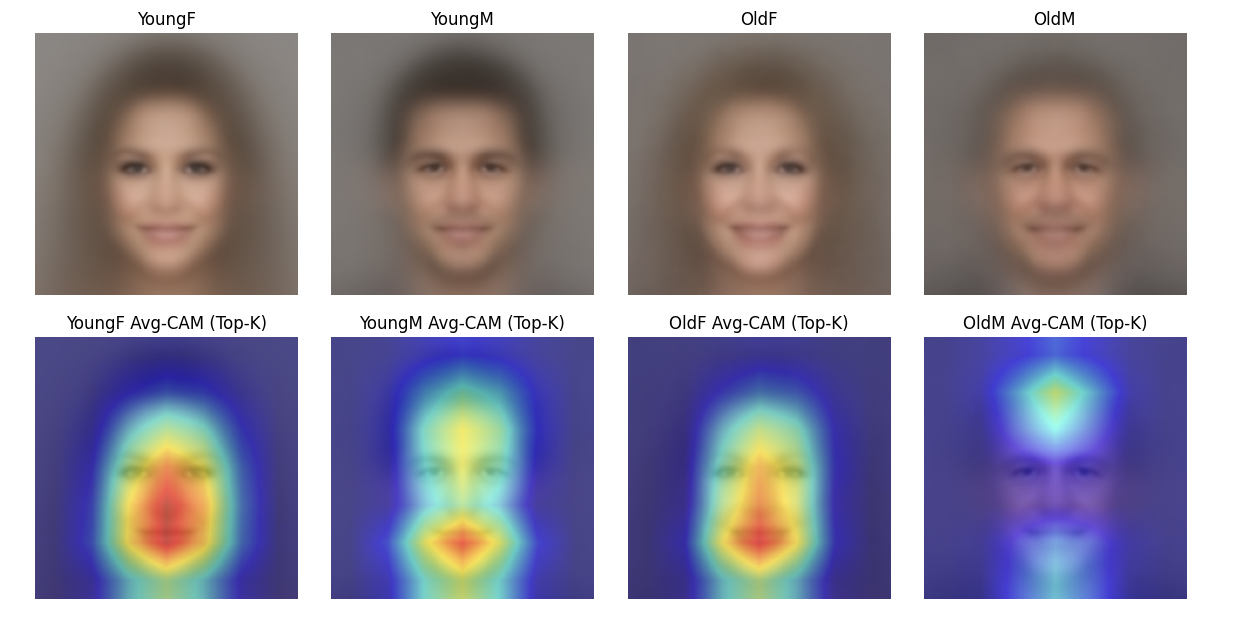}
  \caption{Average facial composites (top) and Grad-CAM overlays (bottom) for young female-perceived, young male-perceived, older female-perceived, and older male-perceived faces, from left to right. Grad-CAM maps for the positive ``Attractive'' output are averaged across the 64 highest-scoring images in each subgroup. Comparisons concern spatial localisation rather than absolute activation strength across groups.}
  \label{fig:grad-cam}
\end{figure}

The averaged Grad-CAM maps show qualitative differences in spatial localisation among the 64 highest-scoring images in each subgroup (Figure~\ref{fig:grad-cam}). Young female-perceived faces show activation across the central face, particularly around the nose and mouth. Older female-perceived faces show a similar central pattern extending along the nose toward the mouth. Young male-perceived faces show two prominent regions around the forehead and mouth, while older male-perceived faces show predominantly upper-forehead activation, with weaker activation near the lower face and neck.

These visualisations describe localisation within the selected high-scoring images. They do not identify the specific appearance properties used within each region or explain the subgroup differences in AP. Their role is to complement the annotation and feature-attribution analyses by showing where spatial sensitivity differs across subgroups.

\section{Discussion}
This study examines gendered appearance double standards as
one manifestation of representational bias. Its central insight
is that favourable representation and unequal evaluation can
coexist: female-perceived faces receive the ``Attractive''
annotation more frequently overall, yet several appearance
bundles have more negative contrasts with the youthful,
feminine-coded reference bundle for this group. This extends
attention from who receives a favourable label to the
appearance associations within that label. The contribution
is therefore not simply another demographic disparity, but
an empirical examination of how a shared evaluative category
can encode different relationships between appearance and
positive judgment.

\paragraph{From gendered representation to gendered evaluation.}
Media scholarship describes gender as communicated through
recurring combinations of appearance, styling, and presentation
\citep{goffman1979,wolf1991}. CelebA's attribute structure
resembles aspects of these conventions: youth, cosmetics,
adornment, and expression-related cues form one broad bundle,
while facial hair, ageing-related cues, and formalwear cluster
with ``Male.'' This correspondence suggests that cultural
patterns of representation can remain visible in the
attribute structure of a computer vision benchmark. It does
not establish that these traits are intrinsically feminine or
masculine, nor that formalwear measures authority or
professional status.

The evaluative analysis develops this connection further.
Research on gendered ageing argues that youth and appearance
have different implications for the social valuation of women
and men \citep{sontag1972,wolf1991,edstrom2018}. Our findings
are consistent with this asymmetry: the ageing- and
masculine-coded bundle has a substantially more negative
contrast with the youthful, feminine-coded bundle among
female-perceived faces. Importantly, this contrast coexists
with higher overall positivity for female-perceived faces.
An audit restricted to aggregate label frequencies would
therefore leave the gendered structure of these associations
unresolved.

This finding complements prior work showing that image
recognition systems describe women disproportionately through
appearance \citep{schwemmer2020}. Whereas that work examines
which kinds of descriptions groups receive, our analysis
examines how appearance relates to evaluation within one
such category. It also develops the concern identified by
\citet{katzman2023} that evaluative labels can reproduce
social hierarchies through their application. A favourable
label is not necessarily evidence of equitable representation:
its significance also depends on the appearance profiles
with which favourable judgment is associated.

These results nevertheless do not demonstrate that comparable
individuals would receive different judgments solely because
of perceived sex. Images assigned to the same bundle can
differ in their constituent attributes, and the ageing- and
facial-hair-related group contains few female-perceived
images. The sensitivity analysis excluding ``Male'' from
clustering preserves the qualitative pattern but does not
resolve this comparability problem. We therefore interpret
the results as evidence consistent with gendered evaluative
standards within CelebA, rather than causal estimates of
discrimination or a direct measure of the overall range of
acceptable appearances.

\paragraph{Predictive success can preserve contested standards.}
The model analyses connect these annotation patterns to a
second concern: a classifier can successfully predict a
subjective target while reproducing associations that warrant
scrutiny. As \citet{jacobs2021measurement} argue, evaluation
requires examining the relationship between a construct and
its operationalisation. Here, ``Attractive'' is not an
independent criterion against which fairness can simply be
assessed; it is itself a socially situated judgment whose
structure is part of the audit.

XGBoost attributions identify related subgroup-specific
predictive associations. Youth and cosmetic cues generally
contribute positively in the pooled model, while the subgroup
models show contrasting contributions for attributes such as
Chubby and Smiling. These findings indicate that successful
prediction need not involve the same appearance associations
across groups. They extend the annotation analysis by showing
how subgroup-conditioned relationships can also appear in
learned predictions.

However, the relationship between annotation structure and
model attribution is not a one-to-one transfer. Cluster
contrasts compare coverage-defined image groups, whereas
SHAP allocates contributions among features within fitted
models. Separately fitted subgroup models also have different
reference predictions and attribute distributions. The
divergent Pale Skin findings are therefore informative:
they show why agreement across methods cannot be assumed,
and why a single attribution pattern should not be treated
as a direct explanation of a dataset-level contrast.
Although skin tone can carry racialised meanings
\citep{hunter2007,glenn2008}, the binary Pale Skin annotation
and these analyses do not independently establish colorism.

\paragraph{Complementary measurements reveal different concerns.}
The spatial analysis adds an exploratory diagnostic rather
than a further test of the same evaluative contrast.
Grad-CAM localises sensitivity mainly to central facial
regions among selected young female-perceived faces and
to the upper forehead among older male-perceived faces.
This complements attention-based work showing that model
internals can reveal patterns not captured by accuracy
disparities \citep{serianni2025}. However, localisation alone
does not establish which appearance traits are being used
or whether a spatial difference constitutes harm. Because
the maps average the highest-scoring images within each
subgroup, they describe selected examples rather than the
full subgroup distributions. The comparison also varies
both perceived sex and the Young annotation, so it cannot
isolate either dimension.

Likewise, the combination of high accuracy (\(0.9200\))
and low average precision (\(0.3977\)) among older
male-perceived faces shows that performance summaries can
give different impressions of a subgroup. Since AP depends
on positive-label prevalence, this contrast does not
independently establish poorer ranking ability or exclusion
from positive predictions. More fundamentally, neither
accuracy nor AP determines whether the evaluative target
itself encodes appropriate standards.

A related measurement concern arises in language models.
\citet{bai2025explicitly} find stereotypical associations
and discriminatory choices in aligned models that appear
unbiased on conventional evaluations. Although our study
examines a different setting, both highlight the limits of
inferring fairness from reassuring evaluation results.
Here, the additional concern is that accurate prediction
can reproduce group-dependent associations embedded in
the evaluative target itself.

Together, these results illustrate the distinction between
a representational concern and the measurements used to
investigate it \citep{katzman2023}. Annotation contrasts,
feature attributions, spatial sensitivity, and performance
metrics answer different questions. Their value lies in
connecting those questions to a theoretically specified
concern, while making disagreements and evidential limits
visible.

\paragraph{Implications and scope.}

For dataset curation and auditing, the findings support
examining evaluative annotations alongside demographic
coverage and predictive performance. Rebalancing data or
reducing error disparities may leave gendered appearance
associations in labels unexamined. Audits of subjective
targets should distinguish overall label frequencies from
within-group associations, report support for comparisons,
and explain their social significance. The aim is to
scrutinise the standards these labels operationalise,
rather than impose equal attractiveness-label rates.

These patterns do not themselves establish psychological
harm. \citet{wang2025measuring} distinguish pragmatic harms
(changes in beliefs, attitudes, or behaviour) from
experiential harms (negative experiences of encountering
stereotypical outputs). Their image-search experiments
found experiential harm without detecting the predicted
immediate pragmatic effects. Future work could examine
whether encountering attractiveness predictions changes
endorsement of gendered appearance norms or produces
feelings of devaluation and pressure to conform, even
when predictions accurately reproduce dataset annotations.

CelebA's taxonomy also constrains the inquiry. Its binary
``Male'' annotation categorises appearance without recording
self-identification or providing a category for non-binary
identities. Such classifications can reify assumptions
that identity is visually determinate
\citep{keyes2018,hanna2020,baker2020}. This audit examines
judgments within these categories without validating them
as measures of biological sex or gender identity.

Finally, CelebA's celebrity images reflect selection,
styling, and photographic conventions that limit
generalisability. The observed associations may arise
through image selection, attribute definitions, annotation
practices, correlated appearances, and model learning;
our analyses neither separate these processes nor establish
annotator intentions. Future work could compare matched
appearance profiles, account for repeated identities,
and intervene on attribute correlations. Replication across
datasets and cultural contexts would help distinguish
CelebA-specific patterns from broader evaluative standards.

\section{Conclusion}

This study examined whether gendered appearance norms are encoded in CelebA and recur in models trained to predict its “Attractive” annotation. Across dataset structure, feature attribution, and spatial sensitivity, the analyses identified
subgroup-conditioned patterns involving youth, ageing, cosmetic presentation, body-size cues, and expressivity. Higher overall attractiveness-label
rates for female-perceived faces coexisted with stronger negative
contrasts for ageing- and masculine-coded bundles. 

The contribution is to connect socially documented appearance
double standards to measurable patterns in dataset annotations
and model behaviour. Extending audits of demographic disparities
and stereotypical descriptions, this approach examines how
groups are evaluated, including when labels are favourable
and predictions are accurate. The findings support examining
representational bias through complementary measurements,
while distinguishing statistical associations from causal
explanations and evidence of harm. More broadly, dataset
curation and model evaluation should scrutinise the cultural
assumptions embedded in subjective targets alongside
representation and predictive performance.

\section*{Generative AI Usage Statement}

ChatGPT, Claude, and Claude Code were used to assist with
manuscript drafting and revision, methodological discussion,
and analysis-code development and debugging. The authors
determined the final content and retain full responsibility
for the manuscript, analyses, and references.

\appendix
\section{Appendix}

\subsection{SHAP Attribution Summaries}
\label{app:shap}

\begin{figure}[t]
\centering
\includegraphics[width=0.625\columnwidth]{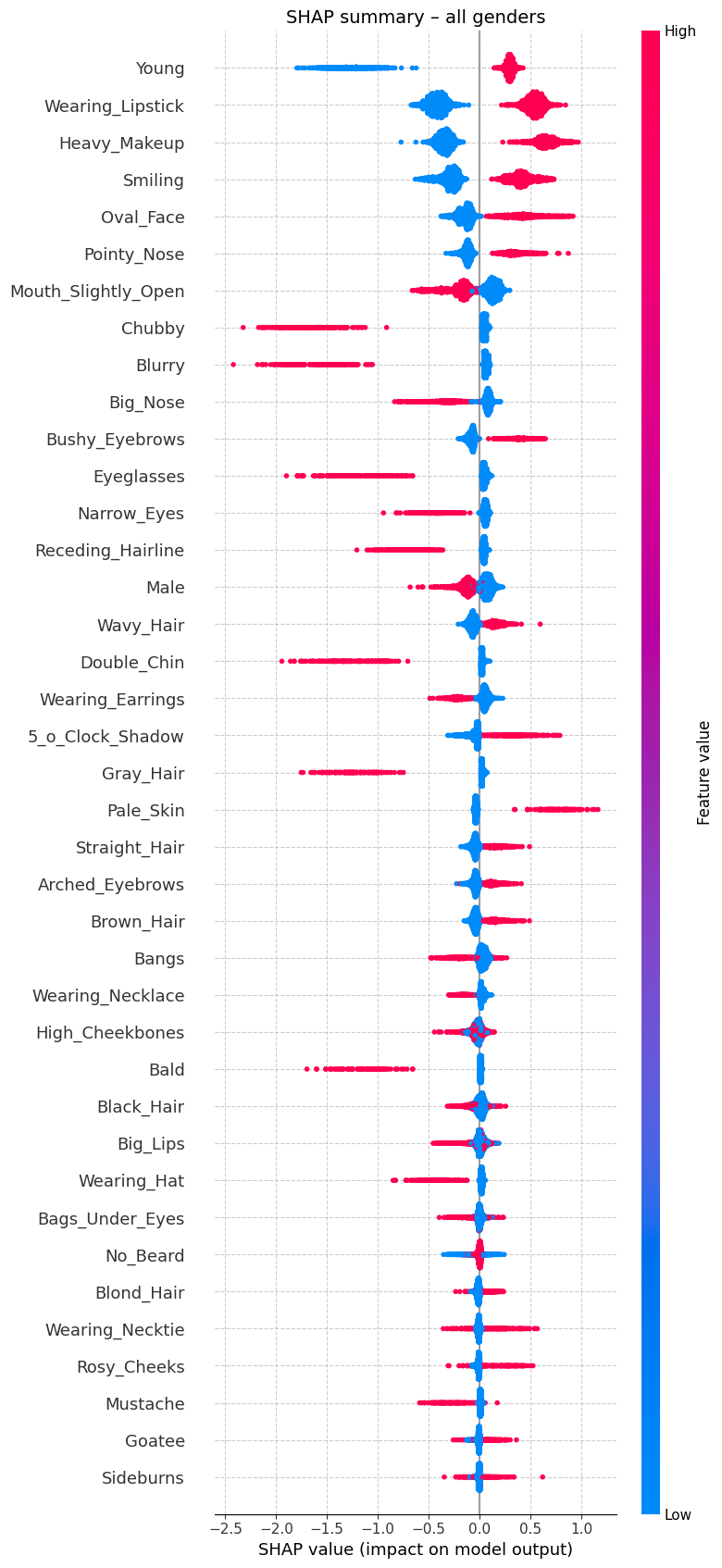}
\caption{Pooled SHAP summary. Horizontal position indicates contribution to model output relative to the reference prediction; colour indicates attribute absence (blue) or presence (red). Features are ordered by mean absolute SHAP value.}
\label{fig:shap_global}
\end{figure}

\begin{figure}[t]
\centering
\includegraphics[width=0.625\columnwidth]{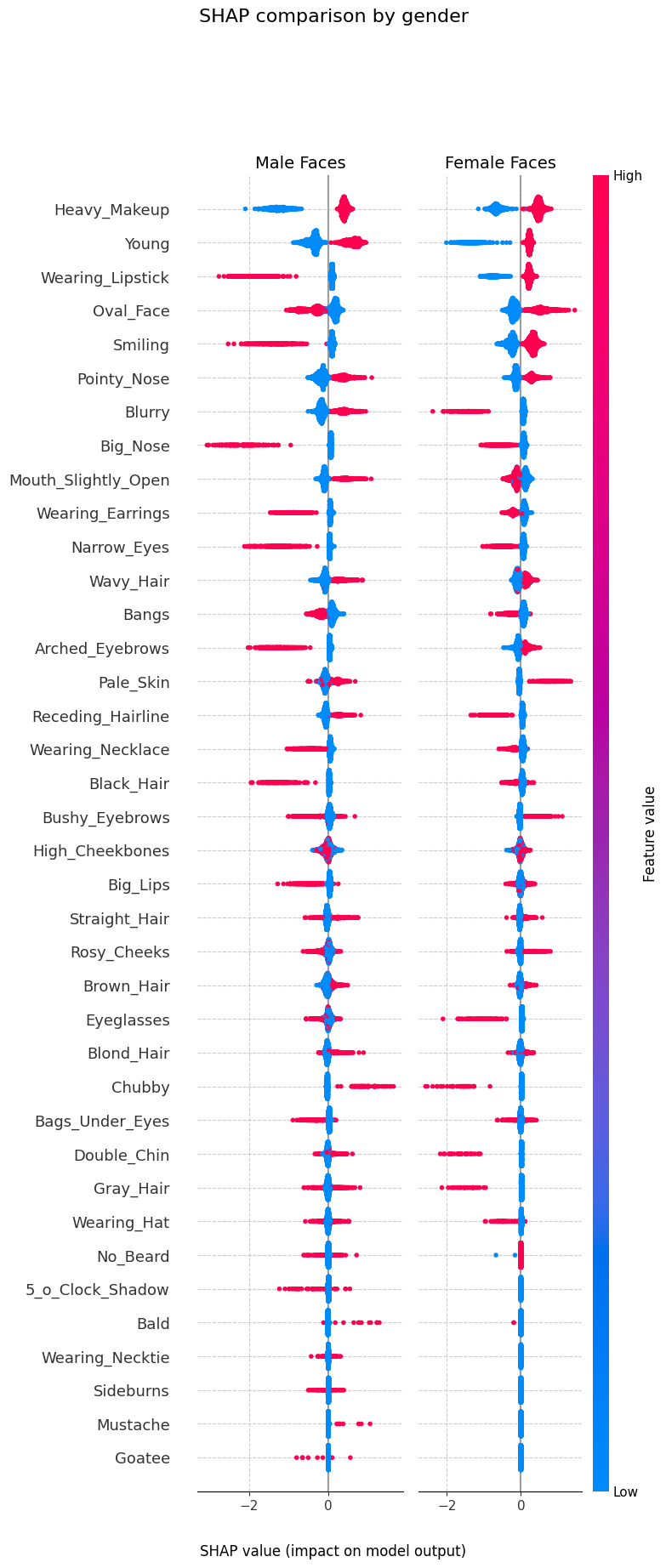}
\caption{SHAP summaries for separate male-perceived (left) and female-perceived (right) classifiers, excluding ``Male.'' Attributions are relative to each model's reference prediction; colour indicates attribute absence (blue) or presence (red).}
\label{fig:shap_gender}
\end{figure}

Figures~\ref{fig:shap_global} and~\ref{fig:shap_gender} present the pooled and subgroup-specific SHAP summaries discussed in the main text.

\subsection{Additional Details on Trait Clustering}

For hierarchical clustering, Pearson correlations between attributes were transformed into dissimilarities using Equation~\eqref{eq:1}. This transformation maps correlations to $[0,1]$, where $0$ indicates perfect positive correlation and $1$ indicates perfect negative correlation.

Hierarchical agglomerative clustering (HAC) was performed using average linkage, which merges clusters according to the mean pairwise dissimilarity between their members. In the resulting dendrogram, leaf nodes represent attributes and branch heights indicate linkage dissimilarity. The correlation matrix was reordered according to the dendrogram's leaf order to produce the clustered heatmap reported in the main text.

\begin{figure}[htbp]
  \centering
  \includegraphics[width=0.8\linewidth]{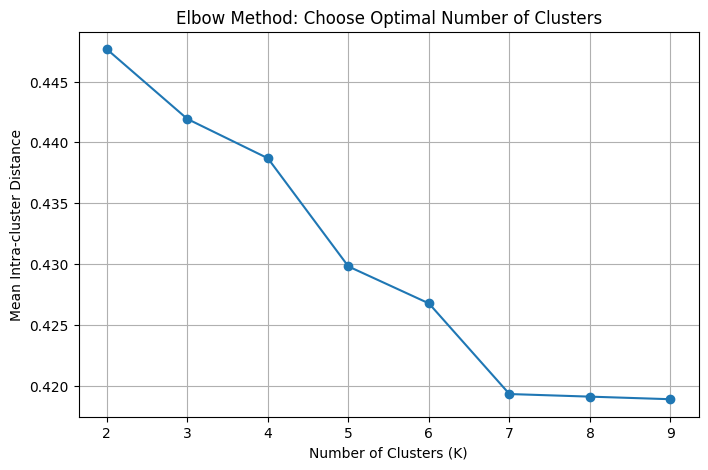}
  \caption{Mean intra-cluster dissimilarity for $K=2$--$9$. The curve flattens beyond $K=7$, indicating diminishing improvements in compactness.}
  \label{fig:elbow}
\end{figure}

The number of clusters was selected using an elbow analysis of mean intra-cluster dissimilarity for $K \in \{2,\ldots,9\}$. The curve decreases substantially up to $K=7$, after which improvements diminish (Figure~\ref{fig:elbow}). This pattern motivated a seven-cluster solution as a practical balance between compactness and interpretability, rather than establishing a uniquely optimal partition.

\begin{figure*}[t]
  \centering
  \includegraphics[width=\textwidth,height=0.70\textheight,keepaspectratio]{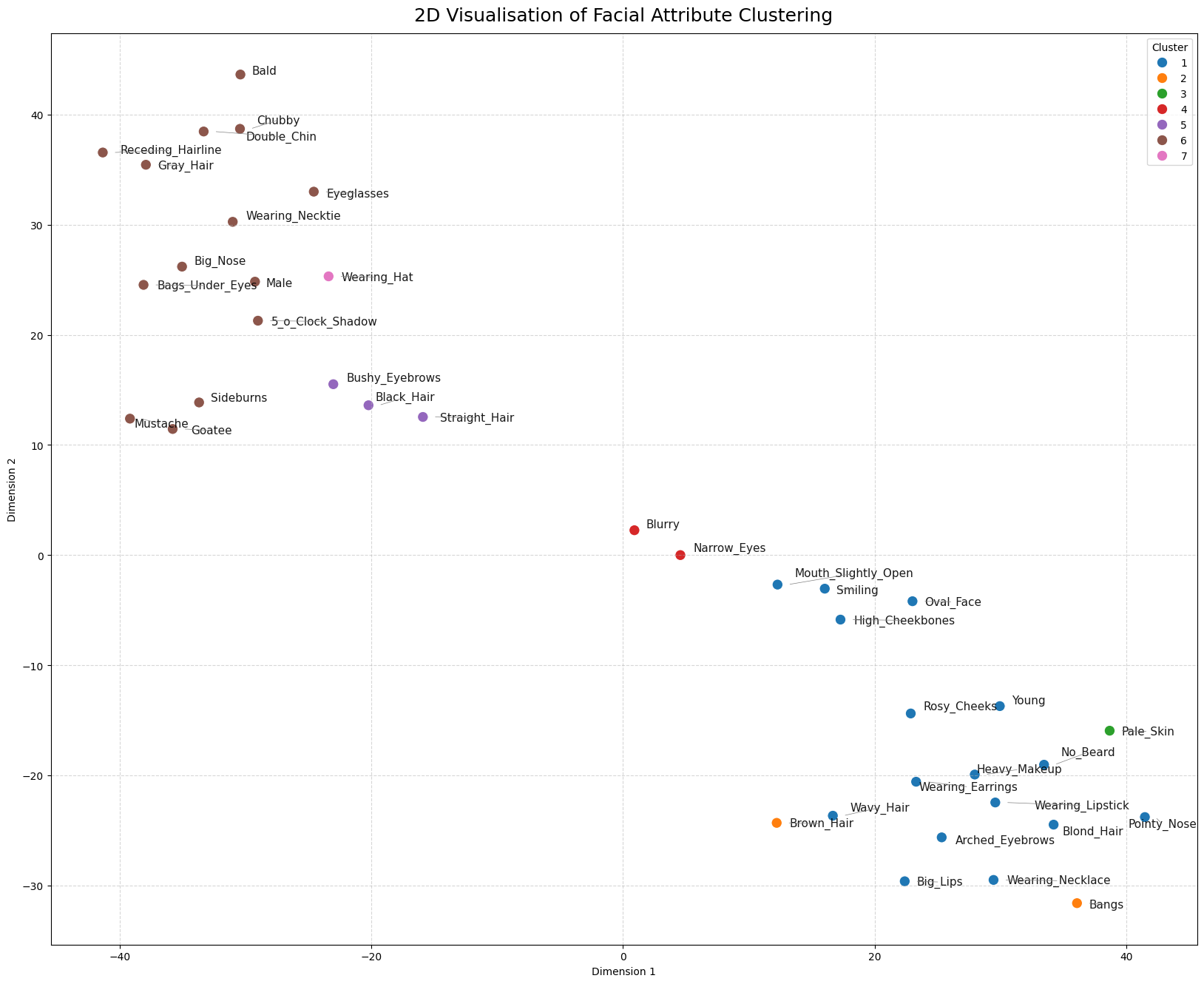}
  \caption{t-SNE projection of the 39 non-target attributes using correlation-based dissimilarities. Colours indicate membership in the seven-cluster HAC solution.}
  \label{fig:tsne}
\end{figure*}

A t-SNE projection of the same dissimilarity matrix provides a complementary visualisation of local attribute relationships (Figure~\ref{fig:tsne}). It is used as an exploratory comparison with the hierarchical structure, rather than independent validation of the clusters or their cultural interpretation.

\subsection{Collinearity Diagnostics and Sensitivity Analysis}

Variance inflation factors (VIFs) were examined to assess collinearity in the logistic regression design matrix (Table~\ref{tab:vif}). Most terms had VIFs below 10, whereas Cluster~6 and its interaction with ``Male'' had VIFs of 136.40 and 138.46, respectively. These values indicate strong dependence between the corresponding model terms. The inclusion of ``Male'' in the cluster's construction and the limited representation of female-perceived faces may contribute to this dependence.

We conducted two sensitivity analyses. First, we repeated clustering and image assignment after excluding ``Male'' from the clustering inputs, while retaining it in the regression and interaction terms. The corresponding ageing- and facial-hair-related bundle was labelled Cluster~2 in the revised solution. Its indicator and interaction with ``Male'' had VIFs of 76.19 and 78.29, respectively. Although lower than in the original specification, these values still indicate substantial dependence. Only 255 of the 21,030 images assigned to this revised cluster were female-perceived, showing that the imbalance persisted without direct inclusion of the demographic label.

The qualitative evaluation pattern also persisted. Relative to the youth- and cosmetics-related bundle, the revised ageing- and facial-hair-related bundle had a more negative association with attractiveness annotations for female-perceived faces on both the probability and log-odds scales. However, the revised bundle differed in attribute composition, so this analysis supports qualitative stability rather than equivalence between the two specifications.

Second, we refitted the original regression without the Cluster~6 indicator and its interaction with ``Male,'' retaining all observations. The remaining coefficients retained their directions and statistical significance, and the perceived-sex contrast within the reference category remained similar. This specification constrains Cluster~6 to share the reference cluster's fitted relationship; it does not exclude Cluster~6 observations. It therefore assesses the sensitivity of the remaining estimates to this constraint, rather than independently validating the Cluster~6 contrast.

Together, these analyses indicate that the qualitative pattern is not solely dependent on directly including ``Male'' in clustering. Nevertheless, estimates involving the ageing- and facial-hair-related bundle warrant caution because of sparse female-perceived representation, uncertainty in the corresponding contrasts, and potentially different attribute compositions within the bundle across perceived-sex groups.

\begin{table}[htbp]
\centering
\caption{Variance inflation factors for the logistic regression design matrix.}
\label{tab:vif}
\begin{tabular}{lc}
\toprule
Predictor & VIF \\
\midrule
Intercept              & 2.89 \\
Cluster 2              & 1.77 \\
Cluster 3              & 1.36 \\
Cluster 4              & 2.36 \\
Cluster 5              & 3.38 \\
Cluster 6              & 136.40 \\
Cluster 7              & 3.42 \\
Male                   & 6.74 \\
Cluster 2 $\times$ Male & 3.37 \\
Cluster 3 $\times$ Male & 1.55 \\
Cluster 4 $\times$ Male & 3.22 \\
Cluster 5 $\times$ Male & 6.43 \\
Cluster 6 $\times$ Male & 138.46 \\
Cluster 7 $\times$ Male & 4.22 \\
\bottomrule
\end{tabular}
\end{table}

\bibliographystyle{ACM-Reference-Format}
\bibliography{references}

\end{document}